\documentstyle[preprint,prl,aps]{revtex}
\begin{document}
\draft
\title{Reply to the comment in quant-ph/0609028 on controlled teleportation}
\author{Chui-Ping Yang}
\address{Department of Physics, Carnegie Mellon university, Pittsburgh, PA 15213}
\maketitle

\begin{abstract}
This is to reply to the comment of Kenigsberg and Mor on our previous work
``Efficient many-party controlled teleportation of multi-qubit quantum
information via entanglement''[Phys. Rev. A 70, 022329 (2004)].
\end{abstract}

\date{\today }

\newpage
Kenigsberg and Mor discussed controlled teleportation in their recent paper
[1]. They made some comments on our previous work about ``Efficient
many-party controlled teleportation of multi-qubit quantum information via
entanglement'' [2]. However, we note that their comments are not correct.
The reasons are as follows.

First, the state of Eq. (3) in Ref. [1] 
\[
\ \otimes _{i=1}^m\left| \phi ^{+}\right\rangle _{AB(i)}\otimes \left| \phi
^{+}\right\rangle _{AC}+\otimes _{i=1}^m\left| \phi ^{-}\right\rangle
_{AB(i)}\otimes \left| \psi ^{+}\right\rangle _{AC} 
\]

is different from the state of Eq. (2) in Ref. [2].

Second, they claimed that Alice and Bob can distill the following mixed
state described by Eq. (4) in Ref. [1] 
\[
\ \otimes _{i=1}^m\left| \phi ^{+}\right\rangle \left\langle \phi
^{+}\right| _{AB(i)}\otimes 1_A+\otimes _{i=1}^m\left| \phi
^{-}\right\rangle \left\langle \phi ^{-}\right| _{AB(i)}\otimes 1_A
\]
and thus Alice can teleport $(m-1)$-qubit state to Bob. However, as a matter
of fact, even based on the state of Eq. (3) used in Ref. [1], one cannot
obtain the above mixed state, after tracing over Carlo's qubit. To see this,
let us rewrite the above state of Eq. (3) in Ref. [1] as follows: 
\begin{eqnarray*}
&&\ \otimes _{i=1}^m\left| \phi ^{+}\right\rangle _{AB(i)}\otimes \left|
\phi ^{+}\right\rangle _{AC}+\otimes _{i=1}^m\left| \phi ^{-}\right\rangle
_{AB(i)}\otimes \left| \psi ^{+}\right\rangle _{AC} \\
&=&\frac 1{\sqrt{2}}\left[ \otimes _{i=1}^m\left| \phi ^{+}\right\rangle
_{AB(i)}\otimes \left( \left| 00\right\rangle _{AC}+\left| 11\right\rangle
_{AC}\right) +\otimes _{i=1}^m\left| \phi ^{-}\right\rangle _{AB(i)}\otimes
\left( \left| 01\right\rangle _{AC}+\left| 10\right\rangle _{AC}\right)
\right]  \\
&=&\frac 1{\sqrt{2}}\left[ \left( \otimes _{i=1}^m\left| \phi
^{+}\right\rangle _{AB(i)}\left| 0\right\rangle _A+\otimes _{i=1}^m\left|
\phi ^{-}\right\rangle _{AB(i)}\left| 1\right\rangle _A\right) \otimes
\left| 0\right\rangle _C\right.  \\
&&+\left. \left( \otimes _{i=1}^m\left| \phi ^{+}\right\rangle
_{AB(i)}\left| 1\right\rangle _A+\otimes _{i=1}^m\left| \phi
^{-}\right\rangle _{AB(i)}\left| 0\right\rangle _A\right) \otimes \left|
1\right\rangle _C\right] 
\end{eqnarray*}

One can check that after tracing over Carlo's qubit from the state of Eq.
(3) in Ref. [1], the density operator for the $m$ EPR pairs shared by Alice
and Bob and the additional qubit held by Alice is 
\begin{eqnarray*}
&&\left( \otimes _{i=1}^m\left| \phi ^{+}\right\rangle \left\langle \phi
^{+}\right| _{AB(i)}+\otimes _{i=1}^m\left| \phi ^{-}\right\rangle
\left\langle \phi ^{-}\right| _{AB(i)}\right) \left( \left| 0\right\rangle
_A\left\langle 0\right| +\left| 1\right\rangle _A\left\langle 1\right|
\right)  \\
&&+\left( \otimes _{i=1}^m\left| \phi ^{+}\right\rangle \left\langle \phi
^{-}\right| _{AB(i)}+\otimes _{i=1}^m\left| \phi ^{-}\right\rangle
\left\langle \phi ^{+}\right| _{AB(i)}\right) \left( \left| 0\right\rangle
_A\left\langle 1\right| +\left| 1\right\rangle _A\left\langle 0\right|
\right) ,
\end{eqnarray*}
which is obviously different from the above mixed state described by Eq. (4)
in Ref. [1].

Last, we think that they might doubt the entangled state of Eq. (2) in Ref.
[2], which they wrote as the state of Eq. (3) in Ref. [1] by a mistake.
However, it is easy to check that the mixed state of Eq. (4) in Ref. [1]
still cannot be created from the state of Eq. (2) in Ref. [2], by tracing
over Carlo's qubit (even for the case when a Hadamard gate is performed by
Alice on her GHZ or EPR qubit).

References

[1] D. Kenigsberg and T. Mor, quant-ph/0609028

[2] C. P. Yang, S. I. Chu, and S. Han, Phys. Rev. A 70, 022329 (2004)

\end{document}